# Forsterite amorphisation by ion irradiation: monitoring by infrared spectroscopy

J.R. Brucato[1] G. Strazzulla[2] G. Baratta[2] and L. Colangeli[1]

[1] INAF Osservatorio Astronomico di Capodimonte, via Moiariello 16, 80131 Napoli, Italy
   e-mail: `brucato@na.astro.it, colangeli@na.astro.it`
[2] INAF Osservatorio Astrofisico di Catania, via S.Sofia 78, 95123 Catania, Italy
   e-mail: `gianni@ct.astro.it, gbaratta@ct.astro.it`

**Abstract.**

We present experimental results on crystal–amorphous transition of forsterite ($Mg_2SiO_4$) silicate under ion irradiation. The aim of this work is to study the structural evolution of one of the most abundant crystalline silicates observed in space driven by ion irradiation. To this aim, forsterite films have been sythesised in laboratory and irradiated with low energy (30–60 keV) ion beams. Structural changes during irradiation with $H^+$, $He^+$, $C^+$, and $Ar^{++}$ have been observed and monitored by infrared spectroscopy. The fraction of crystalline forsterite converted into amorphous is a function of the energy deposited by nuclear collision by ions in the target.

Laboratory results indicate that ion irradiation is a mechanism potentially active in space for the amorphisation of silicates. Physical properties obtained in this work can be used to model the evolution of silicate grains during their life cycle from evolved stars, through different interstellar environments and up to be incorporated in Solar System objects.

**Key words.** methods: laboratory – techniques: spectroscopy – ISM: dust, evolution – cosmic rays

## 1. Introduction

During late stages of evolution, stars loose mass in a short time compared to the life they spend in the main sequence. In the gaseous expanding circumstellar envelopes dust condensation occurs. In particular, the envelopes of giant stars belonging to the asymptotic branch (AGB) (Gehrz 1989) and of supernovae (SN) (Jones 1997) are considered the principal sites of cosmic dust formation. Once formed, grains are injected in the interstellar medium (ISM) driven by the stellar winds.

*Send offprint requests to*: J.R. Brucato

Recent detailed observations of AGB M-type stars in the medium and far infrared spectral regions obtained with the ISO satellite revealed series of bands testifying that both amorphous and crystalline silicate phases are present in the envelopes (Kemper et al. 2001; Molster et al. 2002a, 2002b, 2002c). The results showed the presence of two principal crystalline silicate components, forsterite ($Mg_2SiO_4$) and enstatite ($MgSiO_3$). Relevant correlations between silicate grain properties and outflow or disk source geometries were found. In particular, disk–like stars show strong crystalline silicate bands, while outflow geometry is characterised by weak crystalline silicate bands. Amorphous silicates were detected at temperatures higher than those of crystalline ones. Molster et al. (2002c) showed that enstatite and forsterite masses correlate eachother and that enstatite is more abundant than forsterite by a factor 3-4. Fitting analysis by laboratory spectra of the Spectral Energy Distributions (SED) indicates that, for example, in the source OH-127.8+0.0, 80% of silicates is amorphous olivine (($Mg,Fe)_2SiO_4$), 3% is forsterite, 3% enstatite, 4% metallic iron, and 10% crystalline water ice (Kemper et al. 2002). A value of 3% for forsterite and enstatite was also calculated for many AGB stars (Sylvester et al. 1999) and considered typical for these stars (Kemper et al. 2001).

The distribution of crystalline and amorphous silicate grains according to AGB mass-loss rates was derived (Sogawa & Kozasa 1999; Kemper et al. 2001; Suh 2002). In particular, Kemper et al. (2001) showed that the presence of crystalline silicates does not depend on the mass-loss rate. Moreover, they demonstrated that the lack of spectroscopic signatures of crystalline silicates in the ISO spectra does not imply the absence of crystalline dust in the envelope. This result was obtained considering that, if two separate populations of amorphous and crystalline grains are used in the radiative transfer model of circumstellar envelopes, up to 40 % of crystals can be included in the envelope dust without observing any spectroscopic evidence.

Further analyses of AGB stars spectra were performed by Suh (2002) by using a different dust model. An average single grain population for mixed amorphous and crystalline silicate grains was chosen. It was obtained that about 10 to 20% of crystalline silicates are produced in high mass-loss rate AGB stars, but no crystalline silicates were found in low mass-loss rate stars. The discrepancy of these results depends on the formation mechanism chosen to be active in AGB envelopes. In the first case, two grain populations are formed with different temperatures, while in the second case a single amorphous component is formed that subsequently crystallises totally or partially by annealing, reaching a single final temperature.

The model of Sogawa and Kozasa (1999) already confirmed the absence of crystalline silicates in heterogeneous grains condensed in low mass–loss stars ($\leq 3 \cdot 10^{-5} M_\odot yr^{-1}$). However, condensation models of minerals in stellar winds of M-stars indicate that a multicomponent mixture is mainly formed, dominated by olivine and iron grains (Gail and Sedlmayr 1999). The temperature evolution of the condensed grains in the outflow favours the birth of crystalline grains when their radius is $\leq$ 100 nm. For larger grains the temperature is too low to crystallise the materials, giving rise to a further population of crystalline core grains with an external coating of amorphous material.

Thus, if we consider that AGB stars with circumstellar envelopes are expected to be ubiquitous in our Galaxy (Habing 1996) and that they account for 50% of the total stellar mass-loss of the Galaxy, the crystalline silicate component formed in these stars should be observed in the Interstellar medium (ISM). On the contrary, analysis of ISO spectra showed that an upper limit of only few percent of crystals is present in dense and diffuse ISM (Li and Draine 2001; Demyk et al. 1999).

Among various mechanisms that could explain the absence or the non observability of crystalline silicates in ISM (e.g. selective destruction, low production rate or interstellar dust dilution) in this work we experimentally investigate the amorphisation process by interaction with 30–60 keV ions.

Previous laboratory irradiation experiments of forsterite with energetic (1.5 MeV) protons (Day 1977) and of clinoenstatite with 400 keV and 1 MeV helium ions (Jäger et at. 2003), made at fluences similar to those of cosmic rays, did not show any infrared spectroscopic evidence of structural modification of crystals. Complete amorphisation for low energy (4 and 10 keV) helium ions irradiation at similar fluences was observed (Demyk et al. 2001; Carrez et al. 2002).

The crystal–amorphous transition process can be affected significantly by various physical parameters of the impinging ions, as e.g. the mass, charge and the kinetic energy. Thus, further laboratory studies are needed to investigate the amorphisation process of silicates.

In this work, experimental results on structural modifications suffered by crystalline forsterite under ion irradiation are presented. Thin films of forsterite are synthesised in order to monitor *in situ*, by infrared transmission spectroscopy, the crystalline to amorphous transition during irradiation. Processing effects are investigated, the correlation with the energy deposition is studied and the cross–section of nuclear collision which is directly related to amorphisation is derived. In Section 2 we describe the experimental apparatus and procedures. Results are presented in Section 3 and discussed in Section 4. Astrophysical implications of experimental results and concluding remarks are reported in Section 5.

## 2. Experimental

Among various laboratory techniques which give information on the chemical and physical properties of silicates of various nature, the infrared spectroscopy is widely used. The results obtained by this technique can be used directly for comparisons with space observations, once a dust model is applied. Up to now a series of infrared spectra are available in literature for different classes of crystalline silicates where peak positions and intensities were linked to properties of the samples, as morphology, chemical composition and crystallographic structure (e.g. Koike et al. 1993, Mennella et al. 1998, Jäger et al. 1998, Koike et al. 2000, Fabian et al. 2001, Suto et al. 2002, Chihara et al. 2002, Brucato et al. 2002, Colangeli et al. 2003). In the present work the experimental procedure used is based on the possibility to acquire transmission infrared spectra of crystalline silicate samples and to observe their evolution during irradiation process. To this aim, thin films of crystalline forsterite were prepared in laboratory for *in situ* infrared transmission spectroscopy and ion bombardment processing.

## 2.1. Thin silicate film synthesis

Silicate films were prepared at the Cosmic Physics Laboratory of INAF–Osservatorio Astronomico di Capodimonte by using a Nd–YAG solid state pulsed laser. Its fundamental wavelength output is at 1064 nm with a mean energy of 650 mJ. The power laser output is $10^8$ W cm$^{-2}$ per laser pulse. A set of two crystals was used to get II and IV harmonics at 532 and 266 nm, respectively. The energy output was 80 mJ at 266 nm, 120 mJ at 532 nm and 190 mJ at 1064 nm. By an optical set–up, the 266 nm wavelength was selected and directed onto the target. By using a focusing lens, the power density was mantained at $10^8$ W cm$^{-2}$ by compensating for the decreased mean energy. Targets were prepared by using mixtures of periclase (MgO) and quartz (SiO$_2$) at the stochiometric ratio (2:1) of forsterite Mg$_2$SiO$_4$. The mixture was pressed at 10 tons obtaining pellets 13 mm in diameter and few millimetres thick. The target pellet was mounted inside a vaporisation chamber which was designed to be filled by different gases. The presence of a quenching reactive gas affects the chemical composition of the laser ablated sample, while its pressure drives the morphology of the condensed materials. The lower the pressure inside the chamber the smaller is the size of the condensed grains. In previous laser ablation experiments the vaporisation chamber was filled with oxygen at 10 mbar (Brucato et al. 1999, 2002). This was done to maintain the chemical composition of the condensed sample similar to that of the target. Moreover, the presence of oxygen favours the super–saturation of the hot quenched vapour. This is responsible of the formation of grains with sizes following a log normal distribution with average size of few tens of nanometers. In order to produce films thin enough in order to obtain observable transmission IR spectra before and during ion irradiation, the chamber pressure was maintained at $10^{-5}$ mbar. With these experimental conditions the mean free path of the atoms of the plasma plume produced by the laser ablation can be considered infinitely larger than the distance target–sample (3 cm). Thus, the super saturation condition, necessary to condense grains, is not reached along the path of the atoms to the substrate. Films of amorphous silicates are produced after 15 min deposition on silicon wafer substrates. The infrared spectrum of condensed sample, shown in Figure 1, presents two smooth and broad bands at 10.6 and around 20 $\mu$m typical of amorphous silicate.

In order to prepare samples with crystalline structure, the amorphous silicate films were anneled at 900 °C for 1 hour at the pressure of $10^{-6}$ mbar. The annealing temperature is reached at a rate of 23 C min$^{-1}$ and controlled electronically. The transition from the amorphous to crystalline phase is evidenced by the appearance in the infrared spectum of a series of sharp bands (Figure 1), typical of crystalline materials.

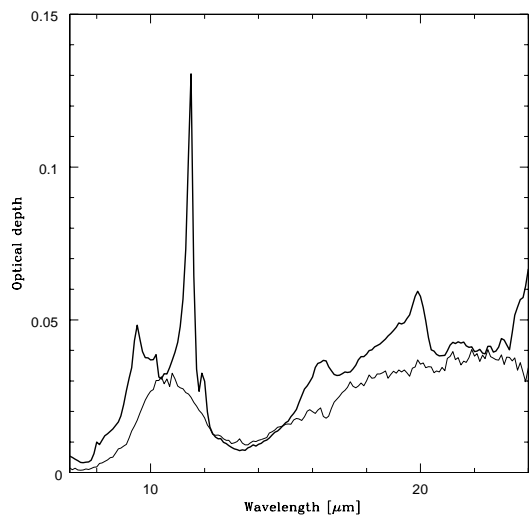

**Fig. 1.** Infrared optical depth spectra of laser synthesised amorphous film (thin line) and of crystalline forsterite after thermal annealing (thick line).

*2.2. Ion irradiation*

Ion irradiation was performed by a Danphysik (1080–30) ion implantation system of INAF–Astrophysical Observatory laboratory of Catania. The gas is ionised in a source and a reduced pressure favours the production of a plasma. Ions are accelerated by 30 kV potential and mean kinetic energies of 30 keV and 60 keV are obtained for once and double ionised ions, respectively. Once extracted, the ions travel a mass separator where defined m/q are selected by a magnetic field. An electrostatic scanning system is used to deflect the ion trajectories. This allows to irradiate uniformly the sample with low current density (of the order of 1 $\mu$A cm$^{-2}$), avoiding undesirable annealing of the sample. The target assembly is designed for *in situ* spectroscopy. For further details on the experimental apparatus see Strazzulla et al. (2001). In the present experiment, samples were irradiated with 30 keV H$^+$, He$^+$, C$^+$ and with 60 keV Ar$^{++}$ with ion fluences up to $10^{17}$ cm$^{-2}$. Light (hydrogen and helium) and heavy (carbon and argon) ions were chosen to study how different classes of ions affect silicate structure.

The thin films were characterised by Fourier transform infrared spectroscopy (Bruker Equinox 55) in the mid-infrared with resolution of 4 cm$^{-1}$. Spectra were acquired after laser deposition and thermal annealing and *in situ* during ion irradiation of samples, in order to monitor the chemical and physical evolution of the silicate films.

## 3. Results

*3.1. Chemical composition and thickness*

In order to check composition and thickness, the IR spectrum of synthesised films was compared with that of crystalline olivine. The optical constants of Mg$_{1.9}$Fe$_{0.1}$SiO$_4$ olivine derived by Fabian et al. (2000) were used to evaluate the optical depth, $\tau$, of the film. This is defined according to the equation $\tau = 4\pi dk/\lambda$, where $d$ is the thickness of the film and $k$ is the immaginary part of complex refractive index. The average assorbance $\tau = \frac{1}{3}[\tau_x + \tau_y + \tau_z]$ along the (x,y,z) crystallographic axes for a film of thickness $d = 380$ Å is reported in Figure 2 and compared with the absorbance of a thin film sample synthesised in this work. Unfortunately, the sample used by Fabian et al. is not the pure forsterite ($Mg_2SiO_4$) end member of olivines, a small fraction of iron being present. The presence of iron in olivine induces a shift in the peak positions towards longer wavelengths that according to Jäger et al (1998) is correlated to the mass percentage of FeO according to the relation: $[FeO]/\Delta\nu = -1.8 \pm 0.1$.

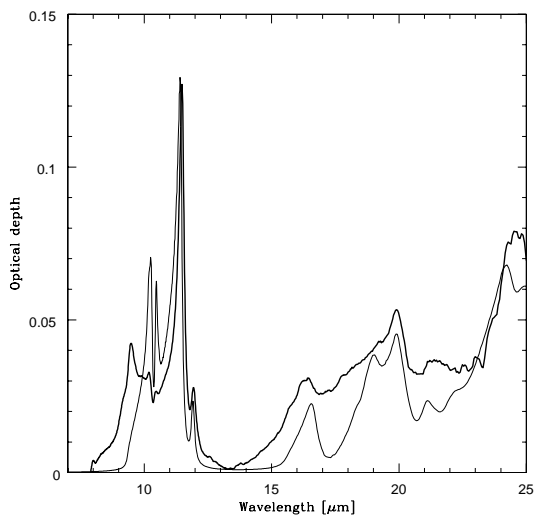

**Fig. 2.** Infrared optical depth spectra of films of synthesised forsterite normalised to the continuum (thick line) and of crystalline olivine ($Mg_{1.9}Fe_{0.1}SiO_4$) calculated by the optical constant of Fabian et al. (2000) (thin line).

Comparing the spectra in Figure 2 it is evident the presence of all the peaks of forsterite in the synthetic sample spectrum, even if a mismatch of peak positions and intensities is observed. Differences in peak intensities are probably due to a preferential axis of growing of the crystals during the thermal annealing. Peak positions are reported in Table 1, together with the peak positions expected for pure forsterite thin film after correction of wavelength shifts. The peak at 9.48 $\mu$m is due to the presence of a fraction of quartz which does not participate to the olivine formation.

In order to give a further estimate of the film thickness, analysis of scanning electron micrographs of the samples was performed. An average film thickness of $500 \pm 200$ Å was obtained. This value is compatible with that obtained by the calculation of the optical depth and comparable with the penetration ranges of the ions (Table 2).

**Table 1.** Peak positions for synthetic forsterite film, olivine $Mg_{1.9}Fe_{0.1}SiO_4$ film as obtained by optical constant from Fabian et al. (2000), and after wavelength shift correction for pure forsterite according to J¨ager et al. (1998).

| Forsterite | Olivine | Forsterite corrected |
|---|---|---|
| 9.48 | – | – |
| 10.19 | 10.25 | 10.28 |
| 10.46 | 10.48 | 10.51 |
| 11.50 | 11.42 | 11.46 |
| 11.94 | 11.92 | 11.96 |
| 16.31 | 16.56 | 16.64 |
| – | 19.02 | 19.12 |
| 19.91 | 19.91 | 20.02 |
| 21.35 | 21.11 | 21.24 |
| – | 22.18 | 22.32 |
| 24.5 | 24.21 | 24.38 |

### 3.2. Ion amorphisation

Infrared spectra of forsterite irradiated with 30 keV $He^+$ at different fluences are shown in Figure 3. A progressive decrease of the intensities of the crystalline peaks is observed increasing the ion fluence and the two large and smooth bands at around 10 and 20 $\mu$m, typical of the amorphous silicates, appear in the spectra. A similar trend is observed in Figure 4, where spectra of forsterite irradiated with 30 keV $H^+$ and $C^+$ are shown. Low intensity residue crystalline peaks are yet observed in Figure 3 for the sample irradiated at the highest fluence. This fluence is that for which an extended ion irradiation does not produce further observable variations in peak intensities. This means that the samples, at the end of the process, are not completely amorphisised. A possible explanation is that the thickness of the film is larger than the ion penetration ranges reported in Table 2. Moreover, ejecta of micron sizes coming from the laser target were observed in scanning electron micrographs, deposited with the film (Figure 5). The inclination of 45 degree of the target surface with respect to the ion beam of the implanter device could prevent a complete ion processing of a fraction of the film which is shielded by the ejecta. Moreover, the ejecta could be forsterite grains with sizes larger than the ion penetration range, that are not completely amorphised by the irradiation.

### 4. Discussion

The spectra in Figures 3 and 4 show that, for about the same ion fluence, the irradiation of forsterite with different ions at the same kinetic energy, produces different amounts of damage of the crystalline structure. This is evidenced by different intensities of sharp (crystalline) and broad (amorphous) bands. In order to quantify the effects on the crystal–amorphous phase transition of forsterite the infrared spectra, $S$, are fitted by a linear combination of crystalline, $S_c$, and amorphous, $S_a$, spectra of forsterite (Figure 1):

$$S = F_c \cdot S_c + F_a \cdot S_a \tag{1}$$

where $F_c$ and $F_a$ are the fractions of crystalline and amorphous components, respectively.

**Table 2.** Nuclear, $S_n$, and electronic, $S_e$, stopping powers and penetration ranges, $R$, in forsterite target computed by TRIM Montecarlo simulation program for ions with kinetic energy, $E$.

| Ion | $E$ [keV] | $S_n$ [eV/Å] | $S_e$ [eV/Å] | $R$ [Å] |
|---|---|---|---|---|
| H$^+$ | 30 | 0.08 | 16 | 2300 |
|  | 1500 | 0.003 | 5.0 | $1.9 \times 10^5$ |
| He$^+$ | 4 | 2.5 | 3.6 | 324 |
|  | 10 | 1.7 | 7.4 | 800 |
|  | 30 | 0.9 | 16 | 2000 |
| C$^+$ | 30 | 12.5 | 27 | 630 |
| Ar$^{++}$ | 60 | 86 | 43 | 431 |

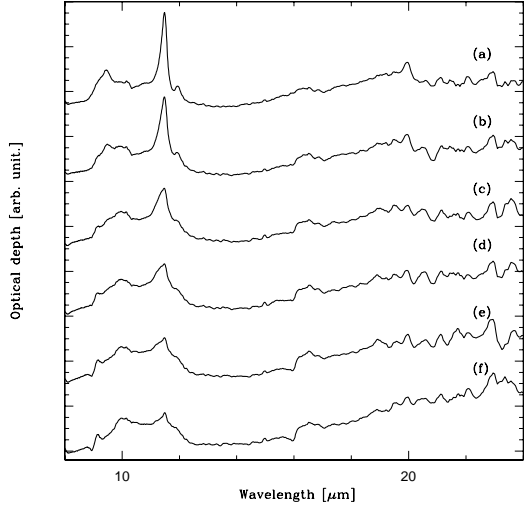

**Fig. 3.** Evolution of infrared optical depth spectra of forsterite (a) before and after irradiation with (b) $8.0 \times 10^{14}$, (c) $4.9 \times 10^{15}$, (d) $1.0 \times 10^{16}$, (e) $2.4 \times 10^{16}$, and (f) $1.04 \times 10^{17}$ He$^+$ cm$^{-2}$ with kinetic energy of 30 keV.

It is also important to evaluate the energy deposition rate on forsterite versus mass, charge, and kinetic energy of the ions. The mean energies deposited by the impinging ions, through nuclear (elastic), $S_n$, and electronic (anelastic), $S_e$, collisions with the target atoms along the ions path were computed by the TRIM Montecarlo simulation program and are reported in Table 2. To evidence if the amorphisation process depends on the energy deposition process in the target, the fractions of crystalline forsterite, $F_c$, remaining after the irradiation versus nuclear, $D_n = \Phi \cdot S_n$, and electronic, $D_e = \Phi \cdot S_e$, irradiation doses are considered, where $\Phi$ is the ion fluence. The results are plotted in Figure 6.

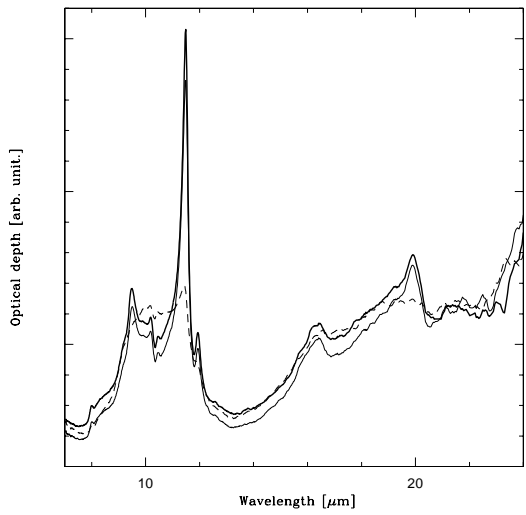

**Fig. 4.** Evolution of infrared optical depth spectra of forsterite before (thick line), after irradiation of 6.8 $10^{15}$ H$^+$ cm$^{-2}$ (thin line), and of 1.4 $10^{15}$ C$^+$ cm$^{-2}$ (dashed line) with kinetic energies of 30 keV. The crystal damage induced by proton irradiation is negligible if compared to that due to carbon ions, even if the kinetic energies are the same and the fluence of proton irradiation is about 5 times larger then that of carbon ions.

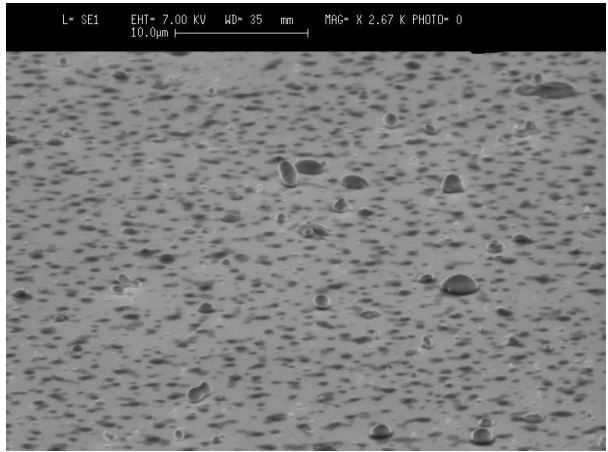

**Fig. 5.** Scanning electron micrograph of forsterite film. Ejecta coming from the laser target are observed with the silicate film.

Similar experiments were performed by irradiating forsterite grains with 1.5 MeV of H$^+$ (Day 1977) and with 4 and 10 keV of He$^+$ (Demyk et al. 2001). The results can be compared with our data in terms of $F_c$ vs. $D_n$. No change in the infrared spectra of H$^+$ irradiated forsterite was observed by Day (1977) ($F_c = 1$), while a complete amorphisation was observed by transmission electron microscopy by Demyk et al. (2001) after irradiation with He$^+$ ($F_c = 0$). Starting from the fluences reported by the authors, $7 \cdot 10^{17}$ 1.5 MeV H$^+$/cm$^2$ (Day 1977) and $5 \cdot 10^{16}$ 4 keV He$^+$/cm$^2$ and $10^{18}$ 10 keV He$^+$/cm$^2$ (Demyk et al. 2001), $D_n$ are calculated by using the $S_n$ of

Table 2. The data well correlate with those obtained in this work, if $F_c$ is reported in funcion of $D_n$, while further missed correlation among the data is evident if $F_c$ is reported in funcion of $D_e$. To quantify the correlation of $F_c$ versus $D_n$, the function:

$$F_c = F_{c0} + A \cdot exp(-\kappa \cdot D_n) \tag{2}$$

is fitted to the data points, where $F_{c0}$ is the asymptotic non irradiated crystalline fraction, $A$ is the fraction of the film which has been amorphised by the ion irradiation and $\kappa$ is the cross–section expressed in cm$^3$ eV$^{-1}$. This quantity represents the volume of crystalline forsterite amorphised per unit energy deposited by elastic collisions. The data points describe the trend of destruction of the crystalline structure and follow well the decreasing exponential law (Figure 6). The best fit is obtained for $F_{c0} = 0.21 \pm 0.03$, $A = 0.74 \pm 0.04$, and $\kappa = 2.2 \pm 0.4 \cdot 10^{-24}$ cm$^3$ eV$^{-1}$ with a coefficient of determination $R^2 = 0.97$. From Figure 6 (bottom panel) it is evident that $F_c$ and $D_e$ do not correlate. The dependence of the forsterite amorphisation on nuclear (elastic) dose demonstrates that it is a consequence of the effect of displacements of the target atoms by nucleus–nucleus collisions due to the impinging ions.

The results obtained in this work confirm that low energy ion irradiation is an efficient process for the amorphisation of silicate grains strongly dependent on $D_n$.

For forsterite, the $S_n$ of the light H, He, and heavy Ar ions were computed by TRIM Montecarlo simulation for kinetic energies in the range 10 eV to 1 GeV (Figure 7). The stopping power increases by increasing the incident ion energy, reaches a maximum and, at higher energies, decreases. The maximum stopping is obtained at higher energies as the mass of the ion increases($S_n$(H)=0.5 eV/Å at 0.28 keV; $S_n$(He)=3.4 eV/Å at 0.7 keV; $S_n$(Ar)=96.2 eV/Å at 20 keV). However, over the whole energy range, the stopping is larger of about 2 orders of magnitude for argon with respect to hydrogen. This means that about 1 % of heavier elements produces the same structural effects than hydrogen on forsterite.

## 5. Astrophysical application

To study if ion irradiation in space is efficient to amorphise forsterite it is necessary to know the flux in space of the ions crossing the grains according to their energies. The intensity of cosmic rays is relatively well known at energies exceeding a few GeV, but becomes increasingly uncertain at lower energies. The intensity measured on Earth understimates the lower–energy particles which are swept back out into the interstellar space due to the presence of the solar magnetic field.

Measures of cosmic rays intensities for nuclei of various charges were made by different authors and extrapolated down to energies as low as few tens of MeV per nucleon. The minimum cosmic ray flux in interstellar space was evaluated by Spitzer and Tomasko (1968) and converted to proton densities by using the relative abundances of particles given by Webber (1967). The

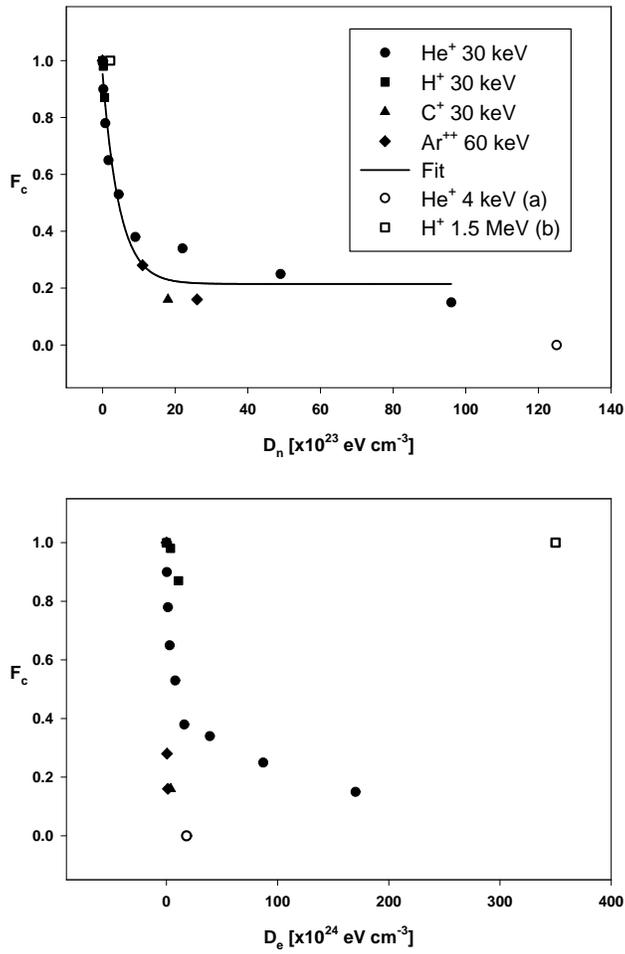

**Fig. 6.** In the top panel the evolution of forsterite crystalline fraction, $F_c$, versus nuclear (elastic) doses, $D_n$ is reported. The thin line represents the fit by the decreasing exponential law $F_c = 0.21 + 0.74 \cdot exp(-2.2 \cdot 10^{-24} \cdot D_n)$ to the evolution of the crystalline fraction of forsterite, $F_c$, versus nuclear dose, $D_n$, for different ion irradiations. The evolution of forsterite crystalline fraction, $F_c$, versus electronic (anelastic) doses, $D_e$, for different ion irradiations in the bottom panel is reported. Values deduced from the experiments of (a) Demyk et al. (2001) and (b) Day (1977) are also reported.

analytical form of the lower limit of cosmic ray intensity adopted by Spitzer and Tomasko (1968) is:

$$J(E) = \frac{0.90}{(0.85 + E_G)^{2.6}} \frac{1}{(1 + 0.01/E_G)} \quad particles \; cm^{-2} \; sec^{-1} \; sterad^{-1} \qquad (3)$$

where $E_G$ is the kinetic energy in unit of $10^9$ eV. This equation is applicable in the range of kinetic energies for nucleon of 10 MeV–10 GeV. Assuming that the time, $t$, spent by a grain of radius $a = 0.1\mu m$ in the interstellar medium is $10^7 - 10^8$ yr, using the cross–section for the amorphisation by ion irradiation of forsterite, derived in this work, the fraction of forsterite grains that is amorphised by high–energy ions can be written as $F_c = 1 - exp(-4\pi^2 a^2 tkJ(E)S_n)$. For proton with energies larger then 10 MeV, $S_n$ is less then $6 \cdot 10^{-4}$ eV Å$^{-1}$ (Figure 7); therefore, the amorphised fraction of the interstellar grains is negligible. Cosmic rays irradiation has no effect on the crystalline infrared features. This is true even if heavy ions induce larger damage on the crystalline structure than hydrogen, according to the different nuclear stopping power (Figure 7).

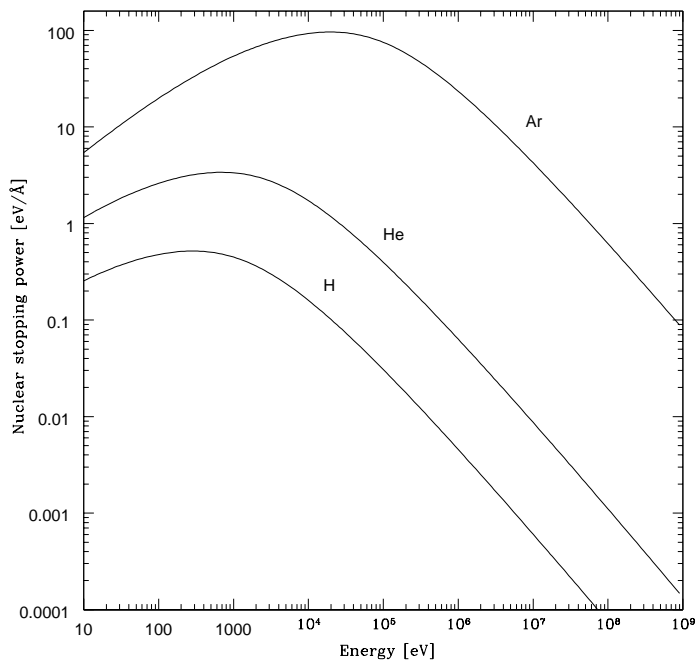

**Fig. 7.** Nuclear stopping power of H, He and Ar ions in forsterite target versus energy, as computed by TRIM Montecarlo simulation program.

Since $S_n$ has its maximum at kinetic energies of the order of tens keV (Figure 7), the effects induced by low–energy ions which are also present in the interstellar space have to be considered. It has been calculated that supernovae generate interstellar shock waves, with velocities $V_s \geq 50$ km s$^{-1}$, able to destroy by sputtering significant amount of dust present in the ISM independently of composition (Seab 1988; Mckee 1989; Jones et al. 1996). High velocity shock waves are able to propagate through large volumes of low density interstellar medium, but are reduced up to vanish in dense molecular clouds (McKee 1989). Moreover, supernova shocks interacting with the interstellar grains may damage the crystalline structures through the mechanism studied in this work as already indicated by Demyk et al. (2001), Carrez et al. (2002), and Jäger et at. (2003). Even if the intensity of low–energy ions as function of their kinetic energy is not known, about $10^{18}$–$10^{19}$ H cm$^{-2}$ are accelerated in a supernova shock at velocities $\geq 100$ km s$^{-1}$ (Jones et al. 1996). These atoms impinge on dust grains with energies $\geq 52$ eV and with $D_n = 4 \cdot 10^{25} - 10^{26}$ eV cm$^{-1}$. According to the data in Figure 6, these doses are sufficient to induce a complete amorphisation of forsterite grains with size of about 40 Å, with consequent large effect on the infrared spectra. This would explaining the broad band silicate spectra observed in the ISM.

The results obtained in this work and those obtained by Demyk et al. (2001), Carrez et al. (2002) and Jäger et at. (2003) demostrate that the amorphisation process due to ion irradiaton may be an efficient process in space and it is able to explain the presence of amorphous silicates in the ISM. In this work it has been shown that the amorphisation process of forsterite dependens on the nuclear elastic collisions between the impinging ions and the target atoms. In contrast, no correlations has been obtained between the irradiation effects on the silicate structure and the anelastic electronic collitions. This result shows that the process of damaging in the case of forsterite acts differently than on water ice, for which a dependence on the total (elastic + anelastic) dose was observed (Leto & Baratta 2003). Large effects on the silicate infrared spectral features are produced by irradiation of low–energy ions that could be responsible of the absence of crystalline peaks in the ISO spectra of silicate dust in the ISM.

*Acknowledgements.* We are grateful to F. Spinella for the technical support given during the ion irradiation experiments. We would like to thank S. Inarta for his collaboration for the SEM analysis. This research has been supported by the MIUR.